\documentclass[prb,preprint]{revtex4-1}
\usepackage{amsmath}  % needed for \tfrac, \bmatrix, etc.
\usepackage{amsfonts} % needed for bold Greek, Fraktur, and blackboard bold
\usepackage{graphicx} % needed for figures
\begin{document}
% Be sure to use the \title, \author, \affiliation, and \abstract macros
% to format your title page.  Don't use lower-level macros to  manually
% adjust the fonts and centering.
\title{Causality in Classical Physics}
% In a long title you can use \\ to force a line break at a certain location.
\author{Asrarul Haque}
\email{ahaque@hyderabad.bits-pilani.ac.in} % optional
\altaffiliation[B209, Department of Physics ]{BITS Pilani Hyderabad
campus, Jawahar Nagar, Shameerpet Mandal, Hyderabad-500078, AP,
India} % optional second address
% If there were a second author at the same address, we would put another
% \author{} statement here.  Don't combine multiple authors in a single
% \author statement.
\affiliation{Department of Physics, BITS Pilani Hyderabad Campus,
 Hyderabad-500078,AP,India}
% Please provide a full mailing address here.
% See the REVTeX documentation for more examples of author and affiliation lists.
\date{\today}
\begin{abstract}
%We address the issue of causality in classical physics.
Classical physics encompasses the study of physical phenomena which
ranges from local (a point) to nonlocal (a region) in  space and/or
time. We discuss the concept of spatial and temporal nonlocality.
However, one of the likely implications pertaining to 'nonlocality'
is non-causality. We study causality in the context of phenomena
involving nonlocality. An appropriate domain of space and time which
preserves causality is identified.
\end{abstract}
% AJP requires an abstract for all regular article submissions.
% Abstracts are optional for submissions to the "Notes and Discussions" section.
\maketitle % title page is now complete
\section{Introduction} % Section titles are automatically converted to all-caps.
% Section numbering is automatic.
Classical physics (Newtonian mechanics and Maxwellian
electrodynamics) deals with the space and/or time varying physical
phenomena of massive point particles and the electromagnetic field.
The physical happenings in classical physics are ordered in time.
What ensures the correct chronological order? It is causality.
Causality, in general, refers to the fact that event $E_1(\vec r,t)$
must occur before in time than event $E_2(\vec r',t'> t)$ if $E_1$
influences $E_2$. For instance, the scalar potential $V(\vec r,t)$
due to an arbitrarily moving point charge reads
\[V(\vec r,t) = \frac{1}{{4\pi \varepsilon _0 }}\int
{\frac{{\rho (\vec r',t - \frac{{|\vec r - \vec r'|}}{c})}}{{|\vec r
- \vec r'|}}} d^3 \vec r';~ (c ~\textup{is the speed of light in
vacuum.)}
\] Charge density $ {\rho (\vec r',t - \frac{{|\vec r - \vec
r'|}}{c})}$ as a cause precedes potential $V(\vec r,t)$ as an
observable effect. The question now arises: how long it takes for
the influence to reach $E_2$ from $E_1$? The \emph{instantaneous}
influence from an event $E_1(\vec r,t)$ to an event $E_2(\vec
r',t')$ is not desirable from the point of view of experience. In
fact, there exist a minimum time $|\vec r'-\vec r|/c$ at which or
only beyond which the disturbance in event $E_1(\vec r,t)$ could be
\emph{sensed} by the event $E_2(\vec r',t')$. Causality, one of the
fundamental principles of physics, requires that
\begin{itemize}
\item the temporal order
of any two events $E_1(\vec r,t)$ and $E_2(\vec r',t')$ must remain
the same for all observers who are moving with constant velocities
with respect to one another and
\item the speed with which interaction can proceed between any two events $E_1(\vec r,t)$ and $E_2(\vec r',t')$
must not exceed the speed of light in vacuum ($3\times 10^8
ms^{-1}$).
\end{itemize}
Therefore no measurable effect can propagate from $(\vec r,t)$ to
$(\vec r',t')$ to surpass the speed of light in vacuum. Unambiguous
distinction between cause and effect in the sense that cause
chronologically precedes effect is thus indispensable to comply the
prediction of a physical theory with the principle of
causality.\\
Classical physics obeys the principle of locality: Newton's equation
of motion
\[
m\frac{{d^2 \vec r}}{{dt^2 }} = \vec F(\vec r,\dot{\vec r},t)
\]
and Maxwell's equations of electromagnetism
\begin{eqnarray}
 \vec \nabla .\vec E(\vec r,t) &=& \frac{{\rho (\vec r,t)}}{{\varepsilon _0 }}
 ;~(\varepsilon _0~ \textup{is electric permittivity in vacuum.}) \\
 \vec \nabla .\vec B(\vec r,t) &=& 0 \\
 \vec \nabla  \times \vec E(\vec r,t)&=&  - \frac{{\partial \vec B(\vec r,t)}}{{\partial t}} \\
 \vec \nabla  \times \vec B(\vec r,t) &=& \mu _0 \vec J(\vec r,t)
 + \frac{1}{{c^2 }}\frac{{\partial \vec E(\vec r,t)}}{{\partial t}};~(\mu _0~ \textup{
 is magnetic permeability in vacuum.)}
 \end{eqnarray}
are both local.
 By locality we mean
 that an event at a given space and time can only influence the events of
 sufficiently \emph{nearby} surroundings.
Locality therefore implies that the state of a particle at time $t$
could be determined by its position $x(t)$ and its velocity $\dot
x(t)$ alone. However, there is a subtle point buried in the
mathematical definition of an \emph{instantaneous} velocity $\dot
x(t)$ with regards to causality. Velocity is defined as
\[
\dot x(t) \equiv \frac{{dx(t)}}{{dt}} \equiv \mathop
{Lt}\limits_{\Delta t \to 0} \frac{{x(t + \Delta t) - x(t)}}{{\Delta
t}}
\]
where ${\Delta t \to 0}$ stands for both ${\Delta t \to 0^+}$ and
${\Delta t \to 0^-}$ as $\Delta t$ can approach to zero from the
right as well as from the left. Let $\Delta t \in ( - \varepsilon
,\varepsilon )$ where $\varepsilon$ is a small positive number. In
the case when ${\Delta t \to 0^+}$, then $x(t + \Delta t)$
corresponds to the positions at later times compared to the position
$x(t)$. The definition of an \emph{instantaneous} velocity now
involves the direction of the movement of position in time from
future to past (as shown in FIG. 1) and thus violates causality.
Whereas for ${\Delta t \to 0^-}$,~$x(t + \Delta t)$ corresponds to
the positions at earlier times with respect to the position $x(t)$.
The movement of position in time in this case runs from past to
future and hence
favors causality.\\
\begin{figure}[hbtp!]
\begin{center}
    \includegraphics[bb = 40 370 500 550, scale=0.7,angle=0]{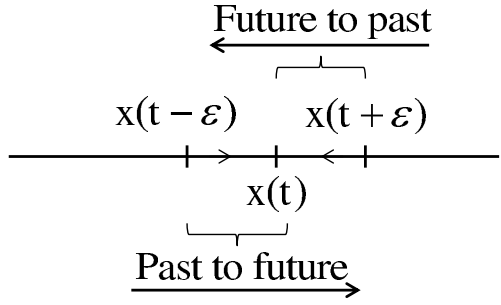}
    \caption{The right-hand derivative $
\mathop {Lt}\limits_{\Delta t \to 0^ +  } \frac{{x(t + \Delta t) -
x(t)}}{{\Delta t}}$ connects future to past whereas the left-hand
derivative $ \mathop {Lt}\limits_{\Delta t \to 0^ -} \frac{{x(t +
\Delta t) - x(t)}}{{\Delta t}}$ connects past to future.}
\label{casa}
\end{center}
\end{figure}
Moreover, classical physics
 deals with the physical phenomena that exhibit nonlocality in the form of
 spread in space and/or time acquired by the physical observable. For instance, for the frequency
 $\omega$ dependent permittivity $\epsilon(\omega)$,
 electric displacement $\vec D$ and electric field $\vec E$ are connected by temporal nonlocality,
\[
\vec D(\vec r,t) = \int\limits_{ - \infty }^{ + \infty }dt'
\epsilon(\vec r,t')\vec E(\vec r,t - t')
\]
in the sense that $\vec D$ has now gotten support over a region
$|t-t'|$ in time through $\vec E$. By nonlocality we mean that an
event in addition to relatively \emph{nearby} events can influence
the suitabley \emph{distant} events as well. Therefore, nonlocality
implies that the state of a particle at time $t$ could be determined
by its position $x(t)$ as well as all \emph{possible} time
derivatives of its position such as ($\dot x(t)$, $\ddot x(t)$,
$\dddot x(t)$,...).
\section{Action-at-a distance}
Interactions in classical physics (Newtonian mechanics) involve
action at a distance. Action at a distance simply means that
interaction could occur between any two distinct spatial points
\emph{instantly}.
 We shall illustrate the concept of action at a distance by
considering two simple systems: one discrete and other
\emph{continuous}. Consider a system of $N$ stationary particles
separated by a distance $R$ each (as shown in FIG.2(a)). Suppose the
particles interact via an action at a distance of range say $R$.
Action at a distance implies that the force between ith and jth
particles
\[
\vec F_{ij} = \vec F_{ij} \left( {\vec r_i (t_i ),\vec r_j (t_j )}
\right)_{t_i  = t_j  = t}
\]
depends on the position of the ith particle as well as jth particle
at the same time. Thus all the particles happen to interact at the
same time. This could be possible provided the interaction
propagates \emph{instantaneously} (i.e. with infinite speed) across
all possible spatial separations in the system. This is what which
conflicts with causality that requires the finite speed of
 propagation of interaction.\\
 Now, consider a particle of mass $m$ moving with velocity $v$ perpendicular to a uniform
 rigid rod collides head-on with its upper edge (please see
FIG.2(b)).
\begin{figure}[hbtp!]
\begin{center}
    \includegraphics[bb = 120 255 500 550, scale=0.7,angle=0]{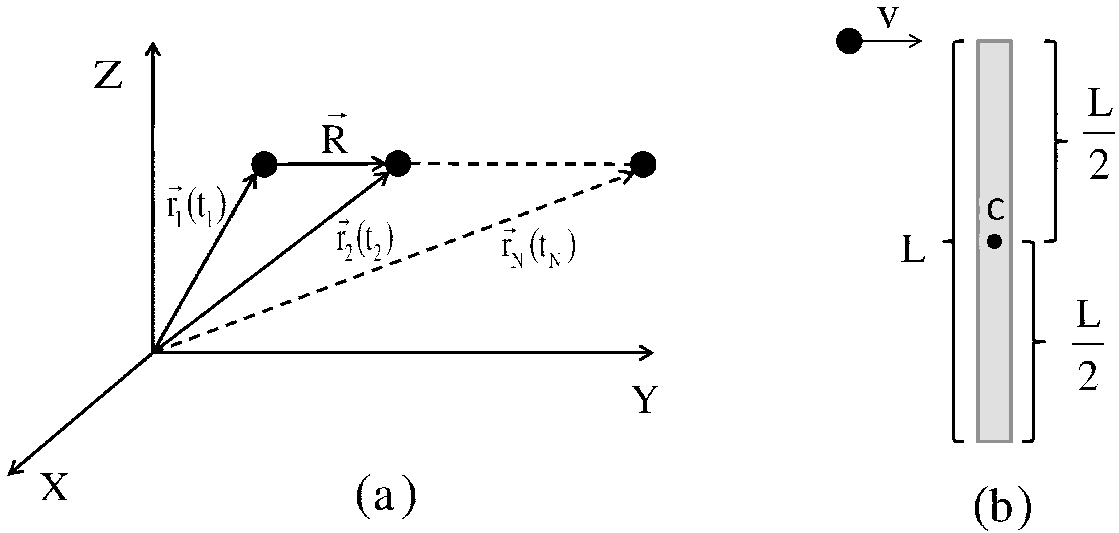}
    \caption{(a) A system of $N$
stationary particles placed at a distance $R$ each interacting via
an action at a distance of range $R$. (b) A particle of mass $m$
moving with velocity $v$ strikes a uniform rod at right angle to the
upper edge.} \label{ad}
\end{center}
\end{figure}
In the calculation of the velocity of the center of mass of the rod,
it is implicitly assumed that the center of mass (in fact every
part) of the rod \textit{senses} the influence of collision at the
same time the body strikes the rod and gets moved
\emph{instantaneously}. However, since the maximum allowable speed
 of the propagation of interaction is $c$, therefore, center of mass would not
\textit{know} as to the collision until a later time $t = L/2c$. The
impulsive force (as a cause) imparted by the particle at the upper
edge and velocity (as an effect) of the center of mass occur at the
same time. Thus
 the identification of cause and effect itself becomes ambiguous.
\section{nonlocality and its side effect}
Locality implies a single space and/or time point support whereas
nonlocality is the smearing of a single space and/or time support.
An observable at a given space and/or time point is said to be
nonlocal if, in addition, it begins to depend upon another space
and/or time point(s). Thus, nonlocality is supported over a region %(size $\sim l$)
 rather than a single space and/or time point. The generic feature of
 nonlocality is the presence of an infinite tower of space and/or time
 derivatives. A nonlocal term such as
 \[
A(s)A(s + l) = A(s)e^{l\frac{d}{{ds}}} A(s)= A(s)\left[ {1 +
l\frac{d}{{ds}} + \frac{1}{2}\left( {l\frac{d}{{ds}}} \right)^2  +
....} \right]A(s)
\]
\[
\begin{array}{l}
 \left[~~ {{\rm{For~ example,~ Let~ }}A(s) = e^s ~\textup{then}~e^{l\frac{d}{{ds}}} e^s
 = \left( {1 + l\frac{d}{{ds}} + \frac{1}{2}\left( {l\frac{d}{{ds}}} \right)^2  + ...} \right)e^s } \right. \\
 \left. { = (1 + l + \frac{1}{2}l^2  + ...)e^s  = e^l e^s  = e^{l + s} =A(s+l)}~~ \right] \\
 \end{array}
\]
possesses an arbitrarily large number of derivatives. Furthermore,
it might involve the propagation of interaction with superluminal
speed (faster than the speed of light) between $s$ and $s+l$
resembling the instantaneous action at a distance over $l$. In
classical physics, nonlocality in space and time generally manifests
in the form of the following expression:
\begin{equation}
\vec A(\vec r,t) = \int\limits_{ - \infty }^{ + \infty } {d^3
r'dt'\vec B(} \vec r - \vec r',t - t')C(\vec r',t') \end{equation}
This relationship is nonlocal in both space and time because $\vec
A$ has now picked up the dependance on the space and time points
other than $ \vec r$ and $t$. $\vec B( \vec r - \vec r',t - t')$ is
an space and time nonlocal function. The above nonlocal relation
becomes local when
\[
\vec B(\vec r-\vec r',t-t') = \vec B_0 \delta (\vec r - \vec
r')\delta (t - t').
\]
Where $\delta (\vec r - \vec r')$ and $\delta (t - t')$ are Dirac
delta functions. The Dirac delta function $\delta (x)(=
\frac{1}{{2\pi }}\int\limits_{ - \infty }^{ + \infty } {e^{ikx} dk}
)$ is defined as follows: $\delta (x)$ is $\infty$ for $x =0$ and
$0$ for $x \ne 0$. The Dirac delta function has a property that $
\int\limits_{ - \infty }^{ + \infty } {f(x)\delta (x - a)dx = f(a)}
$.\\
 $\vec A$ will now turn out to have dependance only on $\vec r$ and $ t$
 as follows
\[
\vec A(\vec r,t) = \int\limits_{ - \infty }^{ + \infty } {dt'} d^3
r'\vec B_0 \delta (\vec r - \vec r')\delta (t - t') C(\vec r',t')
=\vec B_0  C(\vec r,t).
\]
$\vec B$, in general, is however not localized to a point but rather
smeared in space and time. How does the nonlocality in $\vec B$
arise? There exists a transformation called as Fourier
transformation (please see section \ref{ccc}) that can establish a
relation between $\vec B(\vec r-\vec r',t-t')$ and its counterpart
$\vec B(\vec k,\omega )$ in $\vec k$ space and $\omega$ space. We
have,
\begin{eqnarray}
 \vec B(\vec r - \vec r',t - t') &=& \int\limits_{ - \infty }^{ + \infty } {\frac{d^3 k}{(2\pi)^{3/2}}\frac{
 d\omega}{(2\pi)^{1/2}} \vec B(\vec k,\omega )}
  e^{i\vec k.(\vec r - \vec r')} e^{-i\omega (t - t')}  \nonumber\\
  &= &
\vec B\left( { - i\left( {\hat i\frac{\partial }{{\partial x}} +
\hat j\frac{\partial }{{\partial y}} + \hat k\frac{\partial
}{{\partial z}}} \right),i\frac{\partial }{{\partial t}}} \right)
\int\limits_{ - \infty }
  ^{ + \infty } {\frac{d^3 kd\omega}{(2\pi)^2} } e^{i\vec k.(\vec r - \vec r')} e^{-i\omega (t - t')}\nonumber  \\
  &=& (2\pi)^2\vec B( - i\vec \nabla _{\vec r} , i\frac{\partial }
  {{\partial t}})\delta (\vec r - \vec r')\delta (t - t')
 \end{eqnarray}
The \textit{differential operator} $\vec B(-i\vec \nabla _{\vec r} ,
i\frac{\partial }{{\partial t}})$, in the case of nonlocal theory,
involves an arbitrary large number of space and time derivatives and
therefore acting on the product of Dirac delta functions $\delta
(\vec r - \vec r')\delta (t - t') $ will smear them by enlarging
their domain of support. For instance, the \textit{differential
operator} $e^{a\frac{\partial^2}{\partial x^2}}$ acting on
$\delta(x)$ yields
\begin{equation}
 e^{a\frac{{d^2 }}{{dx^2 }}} \delta (x) = e^{a\frac{{d^2 }}{{dx^2 }}} \int\limits_{ - \infty }^{ + \infty } {dk} \frac{{e^{ikx} }}{{2\pi }}%\nonumber \\
  = \int\limits_{ - \infty }^{ + \infty } {dk} e^{ - ak^2 } \frac{{e^{ikx} }}{{2\pi }}
  = \frac{1}{{2\sqrt {\pi a} }}e^{ - \frac{{x^2 }}{4a}}
  \end{equation}
Thus, a single point support $x=0$ is now enlarged to an entire
domain
$(-\infty,\infty)$.\\
One of the foreseeable implications pertaining to nonlocality is
acausality. The immediate reason as to why theories involving
nonlocality are vulnerable to display the symptom of causality
violation could be attributed to the following facts:
\begin{itemize}
\item At a given point in time say $t_0$, interaction can take place between any two spatial
points on the spatial nonlocal region
  exhibiting action at a
spatial distance.
\item At a given point in space say $\vec r_0$, temporal nonlocal connection
could be establish over an entire temporal nonlocal region showing
action at a temporal distance and thus indistinguishing past from
future and viceversa.
\end{itemize}
The following diagrams depict the said facts.
\begin{figure}[hbtp!]
\begin{center}
    \includegraphics[bb = 120 275 500 550, scale=0.7,angle=0]{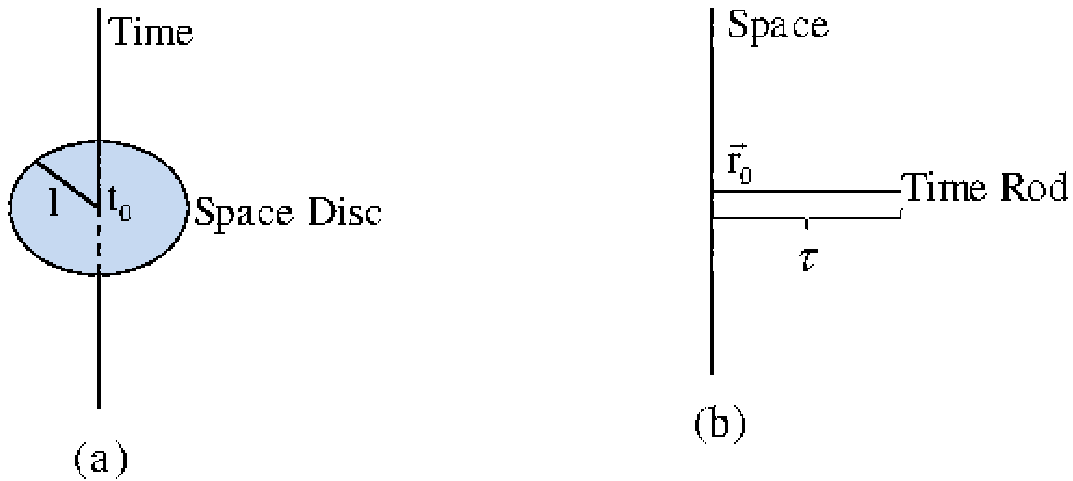}
    \caption{(a) Nonlocal spatial region (space disc) of size $l$ on which any two points can
communicate at a given instant $t_0$. (b) Nonlocal temporal region
(time rod) of size $\tau$ for which any given spatial point say
$\vec r_0$ could exist simultaneously in the past as well as in the
future.} \label{spa}
\end{center}
\end{figure}
\section{Spatio-temporal Nonlocality}\label {ccc}
Let us consider a pair of electric fields oscillating in space and
time as follows:
\begin{eqnarray*}
 \vec E_1  = \vec E_{10} e^{i(\vec k_1 .\vec r - \omega _1 t)}  = \vec E_{10} e^{i\phi _1 (\vec r,t)}  \\
 \vec E_2  = \vec E_{20} e^{i(\vec k_2 .\vec r - \omega _2 t)}  = \vec E_{20} e^{i\phi _2 (\vec r,t)}
% \vec E_3  = \vec E_{30} e^{i(\vec k_3 .\vec r - \omega _3 t)}  = \vec E_{30} e^{i\phi _3 (\vec r,t)}
 \end{eqnarray*}
We shall ask a question, under what condition, the two electric
fields will oscillate with the same wave vector and same frequency?
$\vec E_1 $ and $\vec E_2 $ will oscillate with the same wave vector
provided
\begin{eqnarray*}
 \vec \nabla \phi _1 (\vec r,t)& =& \vec \nabla \phi _2 (\vec r,t)\\
  &\Rightarrow& \vec k_1  = \vec k_2  %= \vec k_3
\end{eqnarray*}
 and the same frequency provided
\begin{eqnarray*}
 \frac{\partial }{{\partial t}}\phi _1 (\vec r,t) &=&
 \frac{\partial }{{\partial t}}\phi _2 (\vec r,t) \\%= \frac{\partial }{{\partial t}}\phi _3 (\vec r,t) \\
  &\Rightarrow& \omega _1  =\omega _2. % = \omega _3.
\end{eqnarray*}
We now consider three physical quantities namely $\vec A$, $B$ and
$\vec C$ that vary in space and time. Suppose these quantities
describe some physical phenomenon in which the first order spatial
and temporal derivatives of their phases happen to be the same. Then
all the three physical quantities will have the same wave vector and
frequency dependance as $\vec A(\vec k,\omega)$, $B(\vec k, \omega)$
and $\vec C(\vec k,\omega)$. We wish to study the causal structure
of the generic form of the ubiquitous equation
\begin{equation}
\vec A(\vec k,\omega) = B(\vec k,\omega )\vec C(\vec k,\omega )
\end{equation}
which embodies spatio-temporal nonlocality in classical physics.
What is the form of spatio-temporal dependence of the above
relation? Point in wave vector/frequecy space corresponds to a
region in the configuration/time space: former is the reciprocal
space of the latter. Such correspondence is established via Fourier
transform which is based on the fact that any \emph{good} function
can be built out of a superposition of sine/cosine functions. The
Fourier transform (FT) of $\vec A(\vec r,t)$ and inverse Fourier
transform (IFT) of $\vec A(\vec k,\omega)$ are defined as follows:
\begin{eqnarray}
 FT\left[ {\vec A(\vec r,t)} \right] &=& \vec A(\vec k,\omega) = \int\limits_{ - \infty }^{ + \infty }
  {\frac{dt}{(2\pi)^{1/2}}} \frac{d^3 r}{(2\pi)^{3/2}}\vec A(\vec r,t)e^{ - i(\vec k.\vec r - \omega t)}  \\
 IFT\left[ {\vec A(\vec k,\omega)} \right] &=& \vec A(\vec r,t) = \int\limits_{ - \infty }^{ + \infty }
  {\frac{d^3k}{(2\pi)^{3/2}} } \frac{d\omega}{(2\pi)^{1/2}} \vec A(\vec k,\omega )e^{i(\vec k.\vec r - \omega t)}
 \end{eqnarray}
Now, $\vec A(\vec r,t)$ can be expressed in terms of $B$ and $\vec
C$ to yield:
\begin{eqnarray}
 \vec A(\vec r,t) &=& \frac{1}{(2\pi)^2}\int\limits_{ - \infty }^{ + \infty } {\vec A(\vec k,t)
 e^{i\vec k.\vec r} e^{i\omega t} } d^3 kd\omega  = \frac{1}{(2\pi)^2}\int\limits_{ - \infty }^
 { + \infty } {B(\vec k,t)\vec C(\vec k,t)e^{i\vec k.\vec r} e^{-i\omega t} } d^3 kd\omega\nonumber  \\
  &= & \frac{1}{(2\pi)^6}\int\limits_{ - \infty }^{ + \infty } {d^3 r'\int\limits_{ - \infty }^
  { + \infty } {d^3 r''} } \int\limits_{ - \infty }^{ + \infty } {dt'\int\limits
  _{ - \infty }^{ + \infty } {dt''} } \int\limits_{ - \infty }^{ + \infty }
  {d^3 kd\omega e^{i\vec k.(\vec r - \vec r' - \vec r'')} e^{i\omega (t' + t''-t )
  } B(\vec r',t')\vec C(\vec r'',t'')}  \nonumber\\
  &= &\int\limits_{ - \infty }^{ + \infty } {d^3 r'\int\limits_{ - \infty }
  ^{ + \infty } {d^3 r''} } \int\limits_{ - \infty }^{ + \infty } {dt'\int
  \limits_{ - \infty }^{ + \infty } {dt''} } \delta (\vec r - \vec r' - \vec
  r'')\delta (t' + t''-t)B(\vec r',t')\vec C(\vec r'',t'')\nonumber \\
  &= &\int\limits_{ - \infty }^{ + \infty } {d^3 r'\int\limits_{ -
  \infty }^{ + \infty }dt'} B(\vec r',t')\vec C(\vec r - \vec r',t - t')
 \end{eqnarray}
 $\vec A(\vec r,t)$ now
depends on the values of $\vec C(\vec r - \vec r',t-t')$ not only at
$\vec r$ and $t$ but in the neighborhood points
 of $\vec r'$ and $t'$ as well. Thus $\vec A(\vec r,t)$ and $C(\vec r - \vec r',t-t')$  are connected by
spatio-temporal nonlocality with $|\vec r- \vec r'|$ and $|t- t'|$
being the scales of spatial and temporal nonlocalities respectively.
We shall now explore the implications of nonlocal connection between
$\vec A$ and $\vec C$. We observe that
\begin{itemize}
\item $\vec A(\vec r,t)$ can depend upon the value of $\vec C(\vec r - \vec r',t - t')$
in the future, i.e. for time $t'>t$ which is possible since for some
given observable time $t$, $t'$ varies from $-\infty$ to $+\infty$.
\item $\vec A(\vec r,t)$ and $\vec C(\vec r - \vec r',t - t')$ could
be connected by a signal which can propagate with speed $\frac{|\vec
r -\vec r'|}{|t-t'|}$ greater than $c$ which seems feasible as
$\frac{|\vec r -\vec r'|}{|t-t'|}$ could be arbitrarily large for
some observable $\vec r$ and $t$ as both $\vec r'$ and $t'$ vary
from $-\infty$ to $+\infty$.
\end{itemize}
\subsection{Spatially nonlocal domain of significance}
Suppose a rapidly changing electric field with an arbitrary space
and time dependance interacts with matter (such as linear
dielectrics). The electric field would create a macroscopic dipole
moment of the system. Suppose the polarization (dipole moment per
unit volume) with the passage of time picks up the same wave number
and frequency dependence as that of the electric field. In linear
dielectrics, the polarization and electric field in wave vector and
frequency spaces are related through the electric susceptibility as
\begin{equation}
\vec P(\vec k,\omega ) = \chi _E (\vec k,\omega )\vec E(\vec
k,\omega )
\end{equation}
which in the configuration and temporal spaces takes the following
form:
\begin{equation}
\vec P(\vec r,t ) = \int\limits_{ - \infty }^{ + \infty } {d^3
r'\int\limits_{ - \infty }^{ + \infty }dt'} \chi_E(\vec r - \vec
r',t - t')\vec E(\vec r',t')
\end{equation}
 When the electric field $\vec E$ possesses a characteristic length scale such as wavelength
($\lambda$) which is comparable to or smaller than the distances
 between the polarization charges then the spatial nonlocal
effects become significant. For $\lambda
 \mathbin{\lower.3ex\hbox{$\buildrel<\over
{\smash{\scriptstyle\sim}\vphantom{_x}}$}}
 l$, where $l$ is the average distance
 between the polarization charges, $\vec E$ can explore the details of the spatial structure of
 dipole moments (see FIG. 4).
Thus, for $\lambda >> l$, spatial nonlocality is not important.
\begin{figure}[hbtp!]
\begin{center}
    \includegraphics[bb = 120 400 500 550, scale=0.7,angle=0]{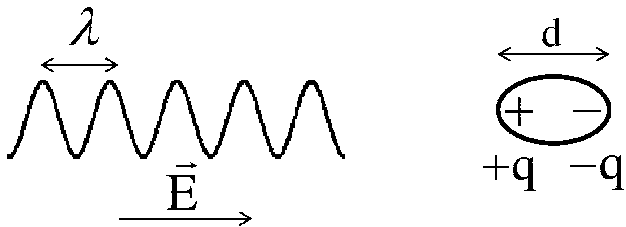}
    \caption{A macroscopic dipole moment of size $d$ created by a sinusoidal varying electric field
    of wavelength $\lambda$.} \label{gasbu}
\end{center}
\end{figure}
\section{Causality}
How the concept of causality is implemented in classical physics? In
classical physics, we often encounter the physical instances where
space and/or time dependent observable effects (such as polarization
$\vec P(\vec r,t)$, steady state displacement $x(t)$ of the forced
oscillator etc.) are caused by space and/or time dependent external
perturbations (such as electric field $\vec E(\vec r,t)$, time
dependent force $F(t)$ etc.). The notion of causality in classical
physics is implemented in the following manner:
\begin{itemize}
\item The physical quantity say $O(\vec r,t)$ which appears as an observable
effect must not contribute anything at time prior to the source
$S(\vec r',t')$ as a cause. Thus,
\begin{equation}
 O(\vec r,t) =0~~~~\textup{for}~~ t<t'
\end{equation}
Therefore, $O(\vec r,t)$ will be nonzero only for $t>t'$.
\item The two space and time points $(\vec r,t)$ and $(\vec r',t')$ can
only be connected by a physical signal provided \begin{equation}
|\vec r-\vec r'|~ < ~c|t-t'|\end{equation} Therefore, $O(\vec r,t)$
must not be different from zero for $|\vec r-\vec r'| > c|t-t'|$.
\end{itemize}
We now study a simple example to understand as to how to impose the
condition of causality. Consider a wave packet [(continuous)
superposition of monochromatic (single wavelength) waves]
$\psi(x,t)$ traveling along the $x$ axis with velocity $v$ as
\[
\psi_I (x,t) = \int\limits_{ - \infty }^{ + \infty }
{\frac{d\omega}{(2\pi)^{1/2}} f(\omega )e^{-i\omega \left(
{\frac{x}{v} - t} \right)} }
\]
Suppose $\psi (x,t)$ is scattered by a particle sitting at the
origin. The scattered wave may be represented as
\[
\psi _S (r,t) = \int\limits_{ - \infty }^{ + \infty }
{\frac{d\omega}{(2\pi)^{1/2}} f(\omega )g(\omega )\frac{e^{-i\omega
\left( {\frac{r}{v} - t} \right)}}{r} }
\]
Suppose $\psi_I (x,t)$ arrives at the scattering center at $t=0$ and
the collision takes place in the neighborhood of time $r/v$. Then
\[
\psi _I (0,t) = 0 ~~~~\textup{for} ~~t<0
\]
with,
\[
f(\omega ) = \frac{1}{{(2\pi)^{1/2} }}\int\limits_0^{ + \infty }
{dt\psi _I (0,t)e^{i\omega t} }
\]
In order to ensure causality for this process, we impose the
condition that the region of integration $t < \frac{r}{v}$ should
not contribute to the scattered wave packet that is
\[\psi _S (r,t)=0~~~\textup{for}~~~t < \frac{r}{v}\]
and \[ f(\omega )g(\omega ) = \frac{1}{{(2\pi)^{1/2}
}}\int\limits_{r/v}^{ + \infty } {dt\psi _S (r,t)e^{ i\omega t} }
\]
Physically this means that the scattered wave packet cannot be
emitted before a time $\frac{r}{v}$ unless the collision between
 the incident wave packet
and the target takes place in the neighborhood of time $r/v$.\\
We now turn to our nonlocal equation. An space and time nonlocal
phenomena in classical physics are generally captured by the
following mathematical relation
\begin{equation}
\label{a1}
 \vec A(\vec r,t) = \int\limits_{ - \infty }^{ + \infty }
 {d^3 r'\int\limits_{ - \infty }^{ + \infty }dt'}  B(\vec r - \vec r',t - t') \vec C(\vec
 r',t').
\end{equation} Suppose $\vec C(\vec r',t')$ serves as a
cause and $\vec A(\vec r,t)$ represents an observable effect in the
above equation. The causality preserving form of equation (\ref{a1})
could be obtained as follows. Let us shift the variable of
integration $\vec r'$ by $\vec \eta \equiv \vec r- \vec r'$ so that
equation (\ref{a1}) becomes
\begin{equation}
\vec A(\vec r,t) =  \int\limits_{ - \infty }^{ + \infty } {dt'}
\int\limits_{ - \infty }^{ + \infty } {d^3 \eta } B(\vec \eta ,t -
t')\vec C(\vec r - \vec \eta ,t')
\end{equation}
where $\vec \eta  = \vec r - \vec r'$. Causality imposes
restrictions on space and time by demanding that $\vec A(\vec r,t)$
will be nonzero provided $t'< t$ and $|\vec\eta| < c|t-t'|$. Thus
\begin{eqnarray}
 \vec A_{causal} (\vec r,t) &=&  \int\limits_{ - \infty }^{ + \infty } {dt'}
\int\limits_{ - \infty }^{ + \infty } {d^3 \eta } B(\vec \eta ,t -
t')\vec C(\vec r - \vec \eta ,t')
 \theta (t - t')\theta \left( {c\left| {t - t'} \right| - \left| {\vec \eta } \right|} \right) \nonumber\\
 & = &  \int\limits_{ - \infty }^t {dt'} \int\limits_0^{2\pi } {d\phi } \int\limits_0^\pi
  {d\theta } \sin \theta \int\limits_0^{c|t - t'|} {d\eta } \eta ^2 B(\vec \eta ,t - t')\vec C(\vec r - \vec \eta ,t')
 \end{eqnarray}
where $\theta (t - t')$ and $\theta \left( {c\left| {t - t'} \right|
- \left| {\vec \eta } \right|} \right)$ are step functions defined
by
\begin{eqnarray*}
 \theta (t - t') &=& 1~~\textup{for}~~t > t'~~ \textup{and}~~0~~\textup{for}~~t < t',\\
 \theta \left( {c\left| {t - t'} \right| - \left| {\vec \eta } \right|} \right)
 &=&
 1~~\textup{for}~~c\left| {t - t'} \right| > \left| {\vec \eta } \right|~~
 \textup{and}~~0~\textup{for}~c\left| {t - t'} \right| < \left| {\vec \eta }
 \right|.
 \end{eqnarray*}
 Moreover, in the case of only temporal nonlocality , we can have
 \begin{equation}
\vec A_{causal}(\vec r,t) = \int\limits_{ - \infty }^{ + \infty }
{dt'} B(\vec r,t - t')\vec C(\vec r,t')\theta (t - t') =
\int\limits_{ - \infty }^t {dt'} B(\vec r,t - t')\vec C(\vec r,t')
\end{equation}
However, for exclusively spatial nonlocality, at a given
\emph{instant} of time, interaction can take place between any two
distinct points belonging to the region of nonlocality and hence
causality is violated in the sense of action at a distance.
\section{Conclusion}
Non-locality is one of the crucial attributes of classical physics.
However, one of the consequences
of the nonlocality might be causality
violation.
Causality in classical physics is ensured by enforcing appropriate constraints
 on space and time. Unambiguous
 distinction between cause and effect demands cause must exist anterior to the effect as well
 as a signal between cause and effect can not connect by superluminal speed. The identification of
causal space and time domain is therefore rather important to
prevent causality violation.

\vspace{0.3in}\
 \textbf{\small
SUGGESTED READING}
\begin{enumerate}
\item J. D. Jackson, \textit{Classical Electrodynamics} (New York:
John Wiley $\&$ Sons, 2003), p.~330-333.
\item H. M. Nussenzveig, \textit{Causality and Dispersion Relations}
(New York: Academic press, 1972), p.~3-16.
\item D.I. Blokhintsev, \textit{Space and Time in the Microworld}
 (Dordrecht-Holland: D. Reidel Publishing Company, 1973), p.~191-199.
\end{enumerate}
\end{document}